%% file: report.tex
\documentclass[a4paper,conference]{IEEEtran}
\usepackage{ifthen}
\newboolean{arxiv}
\setboolean{arxiv}{true}

\usepackage{amsmath,amssymb}
\usepackage{amsthm} % conflict with theorem package
\usepackage{ascmac}
\usepackage{bbm}
\usepackage{bm} % bold math
\usepackage{braket}
\usepackage{cite}
\usepackage{dcolumn} % Align table columns on decimal point
\usepackage{enumerate}
\usepackage{enumitem}
\usepackage{framed}
\usepackage{graphicx} % Include figure files
\ifthenelse{\boolean{arxiv}}{}{\usepackage[dvipdfmx]{hyperref}}
\usepackage{ifthen}
\usepackage{longtable}
\usepackage{mathtools}
\usepackage{multirow}
\usepackage{overpic}
\usepackage{pict2e}
\ifthenelse{\boolean{arxiv}}{}{\usepackage{pxjahyper}}
\usepackage[usestackEOL]{stackengine}
\usepackage{txfonts} % \coloneqq
\usepackage{xparse}

\usepackage{color}
\usepackage[most]{tcolorbox}

\newcommand{\hypercolor}{blue}
\hypersetup{
  colorlinks,
  citecolor=\hypercolor,
  linkcolor=\hypercolor,
  urlcolor=\hypercolor,
  bookmarksopen=true,
  bookmarksopenlevel=4,
  bookmarksnumbered
}
\ifthenelse{\boolean{arxiv}}{}{\mathtoolsset{showonlyrefs,showmanualtags}}

\input{settings_quant.tex}

\newboolean{figure}
\setboolean{figure}{true}

\ifthenelse{\boolean{figure}}{}{}

\newcommand{\relmiddle}[1]{\mathrel{}\middle#1\mathrel{}}

\renewcommand{\c}{\circ}

\newcommand{\Mat}{\mathsf{M}}
\newcommand{\FinProb}{\mathsf{FinProb}}
\newcommand{\NCFinProb}{\mathsf{NCFinProb}}
\newcommand{\FinState}{\mathsf{FinState}}
\newcommand{\fdCAlg}{\mathsf{fdC^*}\mathchar`-\mathsf{Alg}}
\newcommand{\St}{\mathsf{St}}
\newcommand{\Chn}{\mathsf{Chn}}
\newcommand{\BR}{\mathbb{BR}}
\newcommand{\cOne}{\underline{\mathbf{1}}}
\newcommand{\SSh}{\mathrm{S}_\mathrm{Sh}}
\newcommand{\SSe}{\mathrm{S}_\mathrm{Se}}
\newcommand{\SvN}{\mathrm{S}_\mathrm{vN}}
\newcommand{\HBFL}{H_\mathrm{BFL}}
\newcommand{\op}{\mathrm{op}}
\renewcommand{\emph}[1]{\textbf{#1}}

\newcommand{\pagetarget}[2]{%
 \phantomsection%
 \label{#1}%
 \hypertarget{#1}{#2}%
}

\definecolor{gray}{RGB}{128,128,128}

\newcommand{\Discard}[1]{}
\begin{document}
\title{A characterization of von Neumann entropy using functors}
\author{
 \IEEEauthorblockN{Kenji~Nakahira}
 \IEEEauthorblockA{
  Quantum Information Science Research Center, \\
  Quantum ICT Research Institute, Tamagawa University \\
  6-1-1 Tamagawa-gakuen, Machida, Tokyo 194-8610 Japan\\
  {\footnotesize\tt E-mail: nakahira@lab.tamagawa.ac.jp} 
  \vspace*{-2.64ex}}
}

\maketitle

\begin{abstract}
 Baez, Fritz, and Leinster derived a method for characterizing Shannon entropy in classical systems.
 In this method, they considered a functor from a certain category to the monoid of non-negative
 real numbers with addition as a map from measure-preserving functions to non-negative real numbers,
 and derived Shannon entropy by imposing several simple conditions.
 We propose a method for characterizing von Neumann entropy by extending their results to quantum systems.
\end{abstract}

% For peerreview papers, this IEEEtran command inserts a page break and
% creates the second title. It will be ignored for other modes.
\IEEEpeerreviewmaketitle

\section{Introduction} \label{sec:intro}

Von Neumann entropy is a key concept in quantum information theory,
which quantifies the ambiguity in quantum states.
Also, Shannon entropy, an important concept in classical information theory, can be regarded as
von Neumann entropy in classical states.
Baez, Fritz, and Leinster derived Shannon entropy as a quantity that characterizes
measure-preserving functions from classical systems to classical systems \cite{Bae-Fri-Lei-2011}.
Specifically, they showed that if a map from such measure-preserving functions
with probability measures to non-negative real numbers is regarded as a functor in category theory
and satisfies certain properties, it is expressed as the difference of Shannon entropy.
In this paper, we try to derive von Neumann entropy (or Segal entropy)
by extending their result to quantum systems.
Parzygnat has recently extended their result \cite{Par-2021}.
The main difference of our method compared to the one of Ref.~\cite{Par-2021}
is the use of conditions that are considered weaker.
In Refs.~\cite{Bae-Fri-Lei-2011} and \cite{Par-2021}, the discussion was limited to
measure-preserving functions (or their extensions to quantum systems, unital *-homomorphisms),
but in this paper, we consider quantities characterizing any quantum channel.
Although not mentioned in this paper, many different approaches are known for characterizing
Shannon entropy and von Neumann entropy (e.g., \cite{Och-1975,Weh-1978,Bau-2015,Con-2020}).

\section{Previous research}

\subsection{Preliminaries}

Let $\Natural$, $\Real$, and $\Complex$ denote the sets of natural numbers, real numbers,
and complex numbers, respectively.
Also, let $[0,1]$ be the set of real numbers between 0 and 1 inclusive.

In this paper, we refer to a finite-dimensional quantum system with decoherence as a quantum system,
or simply a system.
Any system $A$ can be represented in the following form:
\begin{alignat}{1}
 A = \bigoplus_{i=1}^k A_i, \quad A_i \cong \Mat_{n_i},
 \label{eq:A}
\end{alignat}
where $\Mat_n$ is the set of complex square matrices of order $n$
and $n_i$ is a natural number determined by the subsystem $A_i$.
The set of states (i.e., density operators) of a system $A$ is denoted by $\St_A$,
and the set of channels (i.e., trace-preserving completely positive maps) from a system $A$
to a system $B$ is denoted by $\Chn(A,B)$.

\begin{ex}{}{}
 In the case of $k = 1$, we have $A \cong \Mat_{n_1}$.
 $\St_A$ is isomorphic to the set of density matrices of order $n_1$.
\end{ex}

\begin{ex}{}{}
 When $n_1 = \dots = n_k = 1$, we call a system $A$ a classical system
 and its states classical states.
 We often represent a classical system $A$ as $\Complex^X$, where $X$ is a finite set
 with $|X| = k$ elements.
 A classical system $\Complex^X$ satisfies
 $\Complex^X \cong \bigoplus_{i=1}^{|X|} \Mat_1 \cong \Complex^{|X|}$.
 $\St_{\Complex^X}$ is isomorphic to the set of diagonal density matrices of order $|X|$,
 which is also isomorphic to the set of $|X|$-dimensional non-negative row vectors
 whose sum of components are $1$.
 For example, when $|X| = 2$, we have
 \begin{alignat}{1}
  \St_{\Complex^X} &\cong \left\{
  \begin{bmatrix}
   p_1 & 0 \\
   0 & p_2 \\
  \end{bmatrix}
  \relmiddle| p_1, p_2 \ge 0, ~ p_1 + p_2 = 1 \right\} \nonumber \\
  &\cong \left\{
  \begin{bmatrix}
   p_1 \\
   p_2 \\
  \end{bmatrix}
  \relmiddle| p_1, p_2 \ge 0, ~ p_1 + p_2 = 1 \right\}.
 \end{alignat}
 Each state can also be represented as a collection $\{ p_x \}_{x \in X}$ of non-negative real numbers
 satisfying $\sum_{x \in X} p_x = 1$, which can be regarded as a probability distribution.
 In particular, when $|X| = k = 1$, we often simply write $\Complex$, by abuse of notation.
 Without loss of generality, we may assume $\St_\Complex = \{ 1 \}$.
 There are $|X|$ pure states in the classical system $\Complex^X$, which we will denote by
 $\phi^{X}_x$ $(x \in X)$.
 For example, when $|X| = 2$, there are two pure states represented by $[1,0]^\T$ and $[0,1]^\T$,
 where $^\T$ denotes transposition.
 The state $p$ of the classical system $\Complex^X$ can be expressed in the form
 \begin{alignat}{1}
  p &= \sum_{x \in X} p_x \phi^X_x, \quad
  p_x \ge 0 ~(\forall x \in X), ~\sum_{x \in X} p_x = 1.
 \end{alignat}
\end{ex}

A channel $f$ from a system $A$ to a system $B$ is called pure-to-pure%
\footnote{In Ref.~\cite{Bae-Fri-Lei-2011}, a pure-to-pure channel is called a measurement-preserving function.}
if it maps any pure state to a pure state.
For each system $A$, the map
\begin{alignat}{1}
 \St_A \ni \omega \mapsto \Tr \omega = 1 \in \St_\Complex
\end{alignat}
is the unique channel from $A$ to $\Complex$.
This channel, denoted by $\Tr^A$, is pure-to-pure.

Let $\FinProb$ be the following category:
\begin{itemize}
 \item Its each object is a pair $(\Complex^X,p)$ of a classical system $\Complex^X$
       and its state $p \in \St_{\Complex^X}$.
 \item Its each morphism from an object $(\Complex^X,p)$ to an object $(\Complex^Y,q)$ is
       a pure-to-pure channel $f$ from $\Complex^X$ to $\Complex^Y$ such that $q = f \c p$.
       To indicate that the domain of this morphism is $(\Complex^X,p)$,
       we write $f_p$ instead of $f$.
 \item The composite of its morphisms is the composite of channels as maps, and its each identity morphism
       is the identity channel.
\end{itemize}
Also, let $\BR$ be the following category:
\begin{itemize}
 \item It has a single object.
 \item Its each morphism is a real number.
 \item The composite of morphisms is the sum of real numbers, and its identity morphism is $0$.
\end{itemize}
Let $\BR_+$ be the subcategory of $\BR$ restricted to non-negative numbers as morphisms.

For any two classical systems $\Complex^X$ and $\Complex^Y$, the classical system
$\Complex^{X \sqcup Y}$ (where $\sqcup$ denotes disjoint union) can be considered
as their direct sum.
The state $r$ of $\Complex^{X \sqcup Y}$ can be expressed in the form
\begin{alignat}{1}
 r = \lambda p \oplus (1-\lambda) q &\coloneqq \sum_{x \in X} \lambda p_x \phi^{X \sqcup Y}_x
 + \sum_{y \in Y} (1-\lambda) q_y \phi^{X \sqcup Y}_y
\end{alignat}
using some $p \in \St_{\Complex^X}$, $q \in \St_{\Complex^Y}$, and $\lambda \in [0,1]$.
The set of pure states of $\Complex^{X \sqcup Y}$ is
$\{ \phi^{X \sqcup Y}_x \}_{x \in X} \sqcup \{ \phi^{X \sqcup Y}_y \}_{y \in Y}$.
For any two channels $f \in \Chn(\Complex^X,\Complex^{X'})$ and
$g \in \Chn(\Complex^Y,\Complex^{Y'})$%
\footnote{When we say ``for any $f \in \Chn(\Complex^X,\Complex^{X'})$'', unless otherwise stated,
we assume that $\Complex^X$ and $\Complex^{X'}$ are also arbitrary.},
the channel defined by the map
\begin{alignat}{1}
 \lefteqn{ \St_{\Complex^{X \sqcup Y}} \ni \lambda p \oplus (1-\lambda) q } \nonumber \\
 &\quad \mapsto \lambda (f \c p) \oplus (1-\lambda) (g \c q) \in \St_{\Complex^{X' \sqcup Y'}}
\end{alignat}
is denoted by $f \oplus g \in \Chn(\Complex^{X \sqcup Y},\Complex^{X' \sqcup Y'})$.
If $f$ and $g$ are pure-to-pure, then so is $f \oplus g$. 

A functor $H$ from $\FinProb$ to $\BR$ is said to be continuous if,
for any two classical systems $\Complex^X$ and $\Complex^Y$ and any sequence
$\Natural \ni n \mapsto f^{(n)}_{p^{(n)}} \colon
(\Complex^X,p^{(n)}) \to (\Complex^Y,f^{(n)} \c p^{(n)})$ converging to
$f_p \colon (\Complex^X,p) \to (\Complex^Y,f \c p)$,
the sequence $\Natural \ni n \mapsto H(f^{(n)}_{p^{(n)}})$ converges to $H(f_p)$.

\subsection{Baez, Fritz, and Leinster's theorem}

Baez, Fritz, and Leinster proved the following theorem \cite{Bae-Fri-Lei-2011}
(expressed in the above notation):
\begin{thm}{Baez, Fritz, and Leinster (BFL) \cite{Bae-Fri-Lei-2011}}{BFL}
 Assume that a functor $\HBFL$ from $\FinProb$ to $\BR_+$ satisfies the following conditions:
 \begin{enumerate}[label=\arabic*)]
  \item \pagetarget{cond:BFL_continuous}{}
        \emph{Continuity:} $\HBFL$ is continuous.
  \item \pagetarget{cond:BFL_affine}{}
        \emph{Convex linearity:} For any two pure-to-pure channels
        $f \in \Chn(\Complex^X,\Complex^{X'})$ and
        $g \in \Chn(\Complex^Y,\Complex^{Y'})$ and
        any $p \in \St_{\Complex^X}$, $q \in \St_{\Complex^Y}$, and $\lambda \in [0,1]$,
        $\HBFL((f \oplus g)_{\lambda p \oplus (1-\lambda)q})
        = \lambda \HBFL(f_p) + (1-\lambda) \HBFL(g_q)$ holds.
 \end{enumerate}
 Then, there exists a non-negative real number $c$ such that
 \begin{alignat}{1}
  \HBFL(f_p) &= c(\SSh(p) - \SSh(f \c p)),
  \label{eq:HBFL_to_SSh}
 \end{alignat}
 where $f_p$ is any morphism from $(\Complex^X,p)$ to $(\Complex^Y,f \c p)$ in $\FinProb$,
 and $\SSh(p)$ is the Shannon entropy of $p$,
 i.e., $\SSh(p) \coloneqq - \sum_{x \in X} p_x \log p_x$,
 where we set for convenience $0 \log 0 = 0$.
\end{thm}

\subsection{Parzygnat's theorem}

Parzygnat showed that Segal entropy (or von Neumann entropy) can be derived by extending
BFL's theorem (Theorem~\ref{thm:BFL}) to quantum systems \cite{Par-2021}.
We introduce Parzygnat's results after some preparation.
Note that since it is difficult to concisely state Parzygnat's theorem in a self-contained manner,
this paper omits explanations of some terms
(for details, see \cite{Par-2021} and the references cited therein).

When a system $A$ is expressed in the form of Eq.~\eqref{eq:A},
its state $\omega$ can be expressed in the form
$\omega = \sum_{i=1}^k p_i \omega_i$ $~(\omega_i \in \St_{A_i})$.
When expressed in this way, the Segal entropy, $\SSe(\omega)$, of $\omega$ is defined by
\begin{alignat}{1}
 \SSe(\omega) &\coloneqq \SSe(p) + \sum_{i=1}^k p_i \SvN(\omega_i) \nonumber \\
 &= - \sum_{i=1}^k \Tr (p_i \omega_i \log (p_i \omega_i)),
\end{alignat}
where $\SvN$ is von Neumann entropy, i.e., $\SvN(\rho) \coloneqq - \Tr(\rho \log \rho)$.
The Segal entropy $\SSe(\omega)$ is equal to the von Neumann entropy $\SvN(\tilde{\omega})$
when the state $\omega \in \St_A$ is represented as a density matrix $\tilde{\omega}$
of order $n = \sum_{i=1}^k n_i$.
Therefore, it can be said that there is no substantial difference between Segal entropy and
von Neumann entropy for finite-dimensional systems
(while a noticeable difference arises for infinite-dimensional systems).
Also, for any state $p \in \St_{\Complex^X}$ of a classical system, we have $\SSe(p) = \SSh(p)$.

A finite-dimensional $C^*$-algebra $A$ can be expressed in the form of Eq.~\eqref{eq:A}
\cite{Dor-2018}.
We denote a state of such a $C^*$-algebra $A$ by adding a symbol $^*$ such as $\omega^*$.
Each state $\omega^*$ of $A$ can be expressed as a map
\begin{alignat}{1}
 \omega^* \colon A \ni Q \mapsto \Tr(\omega Q) \in \Complex
 \label{eq:omega_star}
\end{alignat}
using a density operator $\omega \in \St_A$.
For each state $\omega^*$, we write $\omega$ for the density operator satisfying
this equation.
Note that when referring to a state of a $C^*$-algebra $A$, it means a map from $A$ to $\Complex$,
not an element of $\St_A$.

Let $\NCFinProb$ be the following category:
\begin{itemize}
 \item Its each object is a pair $(A,\omega^*)$ of a $C^*$-algebra $A$ and its state $\omega^*$.
 \item Its each morphism from an object $(A,\omega^*)$ to an object $(B,\xi^*)$ is
       a state-preserving unital *-homomorphism $f$ from $B$ to $A$ such that $\xi^* = \omega^* \c f$.
       To indicate that the domain of this morphism is $(A,\omega^*)$,
       we write $f_{\omega^*}$ instead of $f$.
 \item The composite of its morphisms is the composite of maps, and its each identity morphism
       is the identity map.
\end{itemize}

Parzygnat proved the following theorem:
\begin{thm}{Parzygnat \cite{Par-2021}}{Par}
 Assume that a functor $H$ from $\NCFinProb$ to $\BR$ satisfies the following conditions:
 \begin{enumerate}[label=\arabic*)]
  \item \pagetarget{cond:Par_continuous}{}
        $H$ is continuous.
  \item \pagetarget{cond:Par_positive}{}
        $H(!^A_{\omega^*}) \ge 0$ holds for any state $\omega^*$ of a $C^*$-algebra $A$,
        with equality for a pure state,
        where $!^A$ is the unique unital *-homomorphism from $\Complex$ to $A$.
  \item \pagetarget{cond:Par_fibred}{}
        $H$ is a fibred functor from the fibration $\NCFinProb \to \fdCAlg$
        to the fibration $\BR \to \cOne$.
  \item \pagetarget{cond:Par_affine}{}
        $H$ is orthogonally affine.
 \end{enumerate}
 Then, there exists a non-negative real number $c$ such that
 \begin{alignat}{1}
  H(f_{\omega^*}) &= c(\SSe(\omega) - \SSe(\xi))
 \end{alignat}
 holds, where $f_{\omega^*}$ is any morphism from $(A,\omega^*)$ to
 $(B,\xi^* \coloneqq \omega^* \c f)$ in $\NCFinProb$.
\end{thm}

In this theorem, Condition~\ref{cond:Par_fibred} defines a functor $H$ as a (Grothendieck) fibration.
This condition seems to make the relationship that should hold between the fibrations
$\NCFinProb \to \fdCAlg$ and $\BR \to \cOne$ more explicit.
Condition~\ref{cond:Par_continuous} of this theorem and Condition~\ref{cond:BFL_continuous}
of BFL's theorem (i.e., Theorem~\ref{thm:BFL}) are essentially the same%
\footnote{By considering the inclusion functor $I \colon \FinProb^\op \to \NCFinProb$
from the opposite category $\FinProb^\op$ of $\FinProb$ as a subcategory of $\NCFinProb$,
we can regard the functor $\HBFL$ in BFL's theorem as the composite functor
$H I \colon \FinProb^\op \to \BR$.}.
Furthermore, Condition~\ref{cond:Par_positive} is closely related to the condition
of BFL's theorem that $\HBFL$ is a functor from $\FinProb$ to $\BR_+$,
and Condition~\ref{cond:Par_affine} is closely related to Condition~\ref{cond:BFL_affine} of BFL's theorem.

In a categorical sense, the four conditions of Theorem~\ref{thm:Par} seem to be elegant,
but they may seem somewhat too strong intuitively.
In what follows, we try to derive Segal entropy (or von Neumann entropy) from different conditions,
which seem intuitively weaker than the conditions of Theorem~\ref{thm:Par}.

\section{Main theorem}

Let $\FinState$ be the following category:
\begin{itemize}
 \item Its each object is a pair $(A,\omega)$ of a system $A$ and its state $\omega \in \St_A$.
 \item Its each morphism from an object $(A,\omega)$ to an object $(B,\xi)$ is a channel
       $f$ from $A$ to $B$ such that $\xi = f \c \omega$.
       To indicate that the domain of this morphism is $(A,\omega)$,
       we write $f_\omega$ instead of $f$.
 \item The composite of its morphisms is the composite of channels as maps, and its each identity morphism
       is the identity channel.
\end{itemize}
Note that the morphisms of $\FinState$ are not limited to pure-to-pure channels (or unital *-homomorphisms).
$\FinProb$ is a subcategory of $\FinState$.
For any two channels $f \in \Chn(A,B)$ and $g \in \Chn(B,C)$ and any $\omega \in \St_A$,
the composite of morphisms $f_\omega \colon (A,\omega) \to (B,f \c \omega)$
and $g_{f \c \omega} \colon (B,f \c \omega) \to (C,g \c f \c \omega)$ is
\begin{alignat}{1}
 g_{f \c \omega} \c f_\omega = (g \c f)_\omega \colon  (A,\omega) \to (C,g \c f \c \omega).
 \label{eq:composition_FinState}
\end{alignat}
A channel $f$ from a system $A$ to a system $B$ is called left-invertible
(or split mono) if there exists a channel $g$ from $B$ to $A$ such that
$g \c f$ is the identity channel.

As in the case of classical systems, for any two systems $A$ and $B$,
their direct sum $A \oplus B$ can be considered.
Each state $\sigma$ of $A \oplus B$ can be expressed in the form
\begin{alignat}{1}
 \sigma = \lambda \omega \oplus (1-\lambda) \xi
\end{alignat}
using some $\omega \in \St_A$, $\xi \in \St_B$, and $\lambda \in [0,1]$.
The states $\omega \oplus 0$ and $0 \oplus \xi$ with zero operators $0$ are orthogonal.
For any two channels $f \in \Chn(A,A')$ and $g \in \Chn(B,B')$, we write the channel defined by
the map
\begin{alignat}{1}
 \lefteqn{ \St_{A \oplus B} \ni \lambda \omega \oplus (1-\lambda) \xi } \nonumber \\
 &\quad \mapsto \lambda (f \c \omega) \oplus (1-\lambda) (g \c \xi) \in \St_{A' \oplus B'}
\end{alignat}
as $f \oplus g \in \Chn(A \oplus B,A' \oplus B')$.
If $f$ and $g$ are pure-to-pure, then so is $f \oplus g$.

A functor $H$ from $\FinState$ to $\BR$ is said to be continuous if,
for any two systems $A$ and $B$ and any sequence
$\Natural \ni n \mapsto f^{(n)}_{\omega^{(n)}} \colon (A,\omega^{(n)}) \to (B,f^{(n)} \c \omega^{(n)})$
converging to $f_\omega \colon (A,\omega) \to (B,f \c \omega)$,
the sequence $\Natural \ni n \mapsto H(f^{(n)}_{\omega^{(n)}})$ converges to $H(f_\omega)$.

For a given functor $H$ from $\FinState$ to $\BR$, let
\begin{alignat}{1}
 S(\omega) \coloneqq H(\Tr^A_\omega),
 \label{eq:S}
\end{alignat}
where $\Tr^A_\omega \colon (A,\omega) \to (\Complex, \Tr \omega = 1)$ is
the morphism corresponding to the unique channel $\Tr^A$ from $A$ to $\Complex$.

In this paper, we claim that the following theorem holds as an extension of
BFL's theorem to quantum systems:
\begin{thm}{main}{main}
 Assume that a functor $H$ from $\FinState$ to $\BR$ satisfies the following conditions:
 \begin{enumerate}[label=\arabic*)]
  \item \pagetarget{cond:main_continuous}{}
        \emph{Continuity:}
        $H$ is continuous.
  \item \pagetarget{cond:main_affine}{}
        \emph{Convex linearity:}
        For any two pure-to-pure channels $f \in \Chn(A,A')$ and $g \in \Chn(B,B')$
        and any $\omega \in \St_A$, $\xi \in \St_B$, and $\lambda \in [0,1]$,
        $H((f \oplus g)_{\lambda \omega \oplus (1-\lambda)\xi})
        = \lambda H(f_\omega) + (1-\lambda) H(g_\xi)$ holds.
  \item \pagetarget{cond:main_positive}{}
        \emph{Positivity for pure-to-pure channels:}
        $H(f_\omega) \ge 0$ holds for any pure-to-pure channel $f \in \Chn(A,B)$
        and any $\omega \in \St_A$
        (in which case, $f_\omega$ is a morphism from $(A,\omega)$ to $(B,f \c \omega)$),
        with equality for a left-invertible channel $f$.
 \end{enumerate}
 Then, there exists a non-negative real number $c$ such that
 \begin{alignat}{1}
  H(f_\omega) = c(\SSe(\omega) - \SSe(f \c \omega)),
  \label{eq:H_to_SSe}
 \end{alignat}
 where $f_\omega$ is any morphism from $(A,\omega)$ to $(B,f \c \omega)$ in $\FinState$.
\end{thm}

Note that Conditions~\ref{cond:main_affine} and \ref{cond:main_positive} can be
weakened as follows:
\begin{enumerate}[label=\arabic*')]
 \setcounter{enumi}{1}
 \item \pagetarget{cond:main_affine_weak}{}
       \emph{Convex linearity:}
       For any two pure-to-pure channels $f \in \Chn(\Complex^X,\Complex^{X'})$ and
       $g \in \Chn(\Complex^Y,\Complex^{Y'})$ from classical systems to classical systems
       and any $p \in \St_{\Complex^X}$, $q \in \St_{\Complex^Y}$, and $\lambda \in [0,1]$,
       $H((f \oplus g)_{\lambda p \oplus (1-\lambda)q})
       = \lambda H(f_p) + (1-\lambda) H(g_q)$ holds.
 \item \pagetarget{cond:main_positive_weak}{}
       \emph{Positivity for pure-to-pure channels:}
       $H(f_p) \ge 0$ holds for any pure-to-pure channel $f \in \Chn(\Complex^X,B)$
       from a classical system $\Complex^X$ and any $p \in \St_{\Complex^X}$
       (in which case, $f_p$ is a morphism from $(\Complex^X,p)$ to $(B,f \c p)$),
       with equality for a left-invertible channel $f$.
\end{enumerate}

We discuss the relationship between this theorem and BFL's theorem (Theorem~\ref{thm:BFL}).
Let $I$ be the inclusion functor from subcategory $\FinProb$ of $\FinState$ to $\FinState$;
then, we can say that the functor $\HBFL$ in BFL's theorem is the composite
$H I \colon \FinProb \to \BR$.
In this case, from Conditions~\ref{cond:main_continuous} and \ref{cond:main_affine_weak} of
the main theorem, Conditions~\ref{cond:BFL_continuous} and \ref{cond:BFL_affine}
of BFL's theorem can be obtained.
Conditions~\ref{cond:main_continuous} and \ref{cond:main_affine_weak} rephrase the corresponding
conditions of BFL's theorem in the terms of the functor $H$ instead of $\HBFL = H I$.
Also, Condition~\ref{cond:main_positive_weak} of the main theorem corresponds to
the condition in BFL's theorem that $\HBFL$ is a functor from $\FinProb$ to $\BR_+$,
and the former condition is stronger than the latter.
In fact, from Condition~\ref{cond:main_positive_weak} of the main theorem,
it is clear that $\HBFL = H I$ maps any morphism of $\FinProb$ to
a non-negative real number, so $\HBFL$ can be regarded as a functor from $\FinProb$ to $\BR_+$.
Roughly speaking, the main theorem can be said to claim that Segal entropy can be derived
by adding Condition~\ref{cond:main_positive_weak} to BFL's theorem.

Let us supplement on Condition~\ref{cond:main_positive} (or Condition~\ref{cond:main_positive_weak})
of the main theorem.
A pure-to-pure channel $f$ maps pure states to pure states.
Furthermore, it is easily seen that if $f$ is also left-invertible, then
it maps mutually orthogonal pure states to mutually orthogonal pure states.
The latter half of Condition~\ref{cond:main_positive} claims that
for such $f_\omega$, the value of $H$ is zero.
As shown immediately below (see Eq.~\eqref{eq:H_to_S}), since $H$ is a functor,
we have $H(f_\omega) = S(\omega) - S(f \c \omega)$.
Here, if we regard $S(\omega)$ as the ambiguity possessed by the state $\omega$,
then $H(f_\omega)$ is the value obtained by subtracting
the ambiguity possessed by the state $f \c \omega$ from
the ambiguity possessed by the state $\omega$, i.e.,
it can be said to be a value representing how much ambiguity is reduced by the channel $f$.
If ambiguity increases, then $H(f_\omega) < 0$ holds.
Condition~\ref{cond:main_positive} claims that any pure-to-pure channel does not increase
such ambiguity, and that $f$ preserves ambiguity if it is also left-invertible.
Equation~\eqref{eq:H_to_SSe} means that $S(\omega)$ is expressed in the form of $c \SSe(\omega)$.

\section{Proof of the main theorem}

We will now prove the main theorem.
Instead of Conditions~\ref{cond:main_affine} and \ref{cond:main_positive},
we will use Conditions~\ref{cond:main_affine_weak} and \ref{cond:main_positive_weak}.
In the proof, we will use BFL's theorem \cite{Bae-Fri-Lei-2011}.
Note that BFL's theorem is based on the result of Ref.~\cite{Fad-1956}.

Since any channel $f \in \Chn(A,B)$ satisfies $\Tr^B \c f = \Tr^A$,
we have that for any $\omega \in \St_A$,
\begin{alignat}{1}
 H(\Tr^B_{f \c \omega}) + H(f_\omega) &= H(\Tr^B_{f \c \omega} \c f_\omega)
 = H(\Tr^A_\omega),
\end{alignat}
where the first equality follows from the functoriality of $H$, and the second equality follows from
the fact that $\Tr^B_{f \c \omega} \c f_\omega = \Tr^A_\omega$,
which is obtained from Eq.~\eqref{eq:composition_FinState}.
Therefore, from Eq.~\eqref{eq:S}, we have
\begin{alignat}{1}
 H(f_\omega) &= S(\omega) - S(f \c \omega).
 \label{eq:H_to_S}
\end{alignat}

Let $\HBFL \coloneqq H I$, where $I \colon \FinProb \to \FinState$ is the inclusion functor.
Then, as already mentioned, from Conditions~\ref{cond:main_continuous} and \ref{cond:main_affine_weak} of
Theorem~\ref{thm:main}, Conditions~\ref{cond:BFL_continuous} and \ref{cond:BFL_affine}
of BFL's theorem are obtained.
Also, from condition~\ref{cond:main_positive_weak} of Theorem~\ref{thm:main},
it is understood that $\HBFL$ is a functor from $\FinProb$ to $\BR_+$.
Therefore, from BFL's theorem, we obtain Eq.~\eqref{eq:HBFL_to_SSh}.
That is, there exists a non-negative real number $c$ such that,
for any pure-to-pure channel $f \in \Chn(\Complex^X,\Complex^Y)$ and $p \in \St_{\Complex^X}$,
\begin{alignat}{1}
 H(f_p) &= c(\SSh(p) - \SSh(f \c p))
 \label{eq:H_to_SSh}
\end{alignat}
holds.
In particular, considering the case of $f = \Tr^{\Complex^X}$, we obtain
\begin{alignat}{1}
 S(p) &= c \SSh(p), \quad \forall p \in \St_{\Complex^X},
 \label{eq:proof_Sp}
\end{alignat}
where we use $S(p) = H(\Tr^{\Complex^X}_p)$ and
$\SSh(\Tr^{\Complex^X} \c p) = \SSh(1) = 0$.

Let us arbitrarily choose a system $A$ and its state $\omega \in \St_A$
and express $A$ in the form of Eq.~\eqref{eq:A}.
Then, there exists a set of orthogonal pure states $\{ \psi_i \in \St_A \}_{i=1}^n$
with $n \coloneqq \sum_{i=1}^k n_i$ such that $\omega$ is represented by
\begin{alignat}{1}
 \omega &= \sum_{i=1}^n \gamma_i \psi_i, \quad \gamma_i \ge 0, \quad \sum_{i=1}^n \gamma_i = 1.
 \label{eq:proof_omega}
\end{alignat}
In this case, we have
\begin{alignat}{1}
 \SSe(\omega) &= - \sum_{i=1}^n \gamma_i \log \gamma_i.
 \label{eq:proof_omega_S}
\end{alignat}
Consider the channel
\begin{alignat}{1}
 f \colon \St_{\Complex^Z} \ni p \mapsto \sum_{i=1}^n \Tr(\phi^Z_i p) \cdot
 \psi_i \in \St_A
\end{alignat}
from the classical system $\Complex^Z$ with $Z \coloneqq \{1,2,\dots,n\}$ to $A$.
Since $f$ maps each pure state $\phi^Z_i$ $(i \in \{1,\dots,n\})$ of $\Complex^Z$
to $\psi_i$, it is pure-to-pure.
Also, consider the channel
\begin{alignat}{1}
 g \colon \St_A \ni \omega \mapsto \sum_{i=1}^n \Tr(\psi_i \omega) \cdot
 \phi^Z_i \in \St_{\Complex^Z}
\end{alignat}
from $A$ to $\Complex^Z$.
Then, it can be seen that $g \c f$ is the identity channel on $\Complex^Z$.
Therefore, $f$ is left-invertible.
For $\omega$ expressed by Eq.~\eqref{eq:proof_omega}, let
$\gamma \coloneqq g \c \omega = \sum_{i=1}^n \gamma_i \phi^Z_i$.
Since $f \c \gamma = \omega$, we obtain
\begin{alignat}{1}
 H(f_\gamma) &= S(\gamma) - S(\omega) = c\SSh(\gamma) - S(\omega),
\end{alignat}
where the first and second equalities follow from Eqs.~\eqref{eq:H_to_S} and \eqref{eq:proof_Sp},
respectively.
On the other hand, since $H(f_\gamma) = 0$ holds from Condition~\ref{cond:main_positive_weak},
we have
\begin{alignat}{1}
 S(\omega) &= c\SSh(\gamma) = - c \sum_{i=1}^n \gamma_i \log \gamma_i = c \SSe(\omega),
\end{alignat}
where the last equality follows from Eq.~\eqref{eq:proof_omega_S}.
Substituting this into Eq.~\eqref{eq:H_to_S}, we obtain Eq.~\eqref{eq:H_to_SSe},
which completes the proof of the main theorem.

\section{Conclusion}

We have proposed a method to extend BFL's theorem to quantum systems
and characterize Segal entropy (or von Neumann entropy).
Specifically, we showed that if a functor from $\FinState$ to $\BR$ satisfies certain properties,
it can be expressed as a constant multiple of the difference in Segal entropy.

\section*{Acknowledgment}

I am grateful to O. Hirota, M. Sohma, and K. Kato for support.
This work was supported by the Air Force Office of Scientific Research under
award number FA2386-22-1-4056.

\bibliographystyle{myieeetr}

\input{report.bbl}
%\bibliography{quant}

\end{document}

%% file: settings_quant.tex
\usepackage{bm}

\theoremstyle{definition}

\tcbuselibrary{theorems}
\tcbset{
  code={},
  colback=white,
  noparskip,
  before=\par\bigskip\noindent,
  coltitle=black,fonttitle=\upshape\bfseries,
  breakable,frame empty,interior empty,colframe=white,
  description delimiters parenthesis,
  enhanced,theorem style=plain,
  boxsep=0mm,
  leftrule=0mm,rightrule=0mm,
  left=0mm,right=0mm,top=0mm,bottom=0mm,
  before skip=10pt,after skip=10pt,
  posstyle/.style={},
  thmstyle/.style={},
  exstyle/.style={},
}
\newtcbtheorem{postulate}{Postulate}{posstyle}{pos}
\newtcbtheorem{thm}{Theorem}{thmstyle}{thm}
\newtcbtheorem[use counter from=thm]{lemma}{Lemma}{thmstyle}{lemma}
\newtcbtheorem[use counter from=thm]{proposition}{Proposition}{thmstyle}{pro}
\newtcbtheorem[use counter from=thm]{cor}{Corollary}{thmstyle}{cor}
\newtcbtheorem{ex}{Example}{exstyle}{ex}

\newcommand{\Real}{\mathbb{R}}
\newcommand{\Complex}{\mathbb{C}}
\newcommand{\Natural}{\mathbb{N}}

\newcommand{\T}{\mathsf{T}}
\DeclareMathOperator{\Tr}{Tr}

\DeclareMathAlphabet{\mymathbb}{U}{BOONDOX-ds}{m}{n}